\documentclass[10pt]{article}

\usepackage{graphicx,epsfig,booktabs}
\usepackage{times}
\usepackage{amsmath,amssymb}

\def\ecoli{{\em E. coli}}
\def\cheyp{{CheY$_{\mbox{p}}$}}
\def\chea{{CheA}}

\begin{document}

\title{Dynamic receptor team formation can explain the high signal transduction gain
in {\em E. coli}}

\author{R\'eka Albert \footnote{Current address: Department of Physics, Pennsylvania 
State University, University Park, PA 16802},  Yu-wen Chiu and Hans G. Othmer\\
School of Mathematics, University of Minnesota, Minneapolis, MN 55455}

\maketitle

\begin{abstract}
Evolution has provided many organisms with sophisticated sensory
systems that enable them to respond to signals in their environment.
The response frequently involves alteration in the pattern of
movement, either by directed movement, a process called {\em taxis},
or by altering the speed or frequency of turning, which is called {\em
kinesis}.  Chemokinesis has been most thoroughly studied in the
peritrichous bacteria {\it Escherichia coli}, which has 4 helical
flagella distributed over the cell surface (Turner et al. 2000), 
and swims by rotating them (Berg $\&$ Anderson 1973).  

When rotated counterclockwise (CCW) the flagella coalesce into a
propulsive bundle, producing a relatively straight ``run'', 
and when rotated clockwise (CW) they fly apart, resulting in a
``tumble'' which reorients the cell with little translocation 
 (Berg $\&$ Brown 1972, Mcnab $\&$ Orston 1977, Turner et al. 2000).  A
stochastic process generates the runs and tumbles, and in a
chemoeffector gradient runs that carry the cell in a favorable
direction are extended. The cell senses spatial gradients as temporal
changes in receptor occupancy and changes the probability of CCW
rotation (the bias) on a fast time scale, but adaptation returns the
bias to baseline on a slow time scale, enabling the cell to detect and
respond to further concentration changes (Block et al. 1982,
 Segall et al. 1986). The overall structure of the
signal transduction pathways is well-characterized in \ecoli, but
important details are still not understood. Only recently has a source
of gain in the signal transduction network been identified
experimentally, and here we present a mathematical model based on
dynamic assembly of receptor teams that can explain this
observation.
\end{abstract}

 \ecoli\ has five receptor types, but most is known about the
aspartate receptor Tar, which communicates with the flagellar motors
via a phosphorelay sequence involving the CheA, CheY, and CheZ
proteins.  CheA, a kinase, first autophosphorylates and then transfers
its phosphoryl group to CheY.  CCW is the default state in the absence
of \cheyp, which binds to motor proteins and increases CW
rotation. Ligand binding to Tar reduces the autophosphorylation rate
of CheA and the rate of phosphotransfer, and thereby increases the
bias. This is the excitation component of the response.  Bacteria also
adapt to constant stimuli, and this is effected by changes in the
methylation state of Tar.  Tar has four residues that are reversibly
methylated by a methyltransferase, CheR, and demethylated by a
methylesterase, CheB.  CheR activity is unregulated, whereas CheB,
like CheY, is activated by phosphorylation via CheA.  Thus, the
receptor methylation level is regulated by feedback signals from the
signaling complex, which can probably shift between two conformational
states having different rates of CheA autophosphorylation.  Attractant
binding and demethylation shift the equilibrium toward a low activity
state of CheA, and attractant release and methylation shift the
equilibrium toward a high activity state.  These key steps, excitation
via reduction of the autophosphorylation rate of CheA when Tar is
occupied, and adaptation via methylation of Tar, have been
incorporated in mathematical models of signal transduction
(Spiro et al. 1997, Barkai $\&$ Leibler 1997, Morton-Firth et al. 1999).

\ecoli\ can sense and adapt to ligand concentrations that range
over five orders of magnitude (Bourret et al. 1991). In addition,
the transduction pathway from an extracellular ligand to the flagellar
motor is exquisitely sensitive to chemical stimuli. Bacteria can
detect a change in occupancy of the aspartate receptor as little as
$0.1-0.2\%$, corresponding to the binding of one or two ligand molecules per
cell. The gain of the system, calculated as the change in rotational
bias divided by the change in receptor occupancy, was found to be
about $55$ (Segall et al. 1986), and a long-standing question is
what the source of this high sensitivity or gain is.  Three main
sources of gain have been suggested: (i) highly cooperative binding of
\cheyp\ to the motor proteins, (ii) regulation of CheZ activity , and
(iii) indirect activation of many receptors by a ligand-bound
receptor.  However, it is known that the high sensitivity is present
in CheZ mutants (Kim et al. 2001), thereby ruling our the second
possibility. Furthermore, it was shown that in the absence of
cooperativity in signal transduction upstream of the motor, a Hill
coefficient of at least $11$ was needed in the response of the motor to
\cheyp\ to explain the observed gains of $3-6$ (Spiro et al. 1997).
Cluzel {\it at al.} (2000) have confirmed this
prediction experimentally, showing that the apparent Hill coefficient
in the functional dependence of the bias on \cheyp\ is about $10$.
However, this cannot account for all the observed gain, and Sourjik
and Berg (2002) have shown, using fluorescence
resonance energy transfer, that the stage between aspartate binding
and \cheyp\ concentration has an amplification $35$ times greater than
expected. None of the existing models of the
full signal transduction system (Spiro et al. 1997, 
Barkai $\&$ Leibler 1997, Morton-Firth et al. 1999) address this source of gain.

Receptor interaction, either directly via clustering, or indirectly
via an intracellular signal, is a likely source of the upstream
component of the gain. Receptors are normally dimeric, and it has been
suggested that ligand binding affects the spatial
packing of the receptor array
(Levit et al. 1998, Parkinson J. S., 
University of Utah, personal communication, 1999). Recent 
experiments show that transmembrane signaling occurs via receptor
clusters or teams, probably of trimers of dimers
(Ames et al. 2002, Kim et al. 2002). It was previously suggested in
analogy with Ising models that clustering may enhance the sensitivity
at low signals, but it is difficult to obtain both high gain and a
wide dynamic range in models of this type
(Bray et al. 1998). Moreover, while these types of models address
the possibility of cooperative interaction as a mechanism for
generating gain, the nature  of this interaction is not specified and
thus  experimental tests are difficult.  More recently an
abstract model based on the energetics of interactions between
receptors was proposed and analyzed by Mello $\&$ Tu (2003). The model
assumes that each receptor dimer can be in an active or inactive
state, and that transitions between these states are rapid compared to
ligand binding. Thus receptors flicker `on' and `off' between these
states, according to an equilibrium distribution, and ligand binding
biases the proportions in the two states. Parameters can be found so
that the model reproduces existing data, but again there is no
molecular mechanism that can be tested. Our goal here is to provide a
more mechanistically-based description of the origin of high gain.
The model is based on the idea that teams of receptor dimers assemble
and dis-assemble dynamically, and that different types of receptors
can assemble in different types of teams. In our analysis assembly and
dis-assembly may occur on comparable time scales, but a static scheme
in which teams exist for long time periods is a limiting case of the
model.

Before describing the model, we observe that the large gain upstream
of the motor can be qualitatively understood, once the
experimentally-determined activity curves are known.  The output of the
signal transduction network as a function of attractant concentration
has been studied in several recent experiments, both {\it in vitro}
(Li $\&$ Weis 2000, Bornhorst $\&$ Falke 2001, Levit $\&$ Stock 2002) and {\it in vivo}
(Sourjik $\&$ Berg 2002). {\it In vitro} experiments use receptor-CheW-CheA
complexes reconstituted in the presence of attractant and measure the
CheA activity immediately after the addition of ATP
(Li $\&$ Weis 2000, Bornhorst $\&$ Falke 2001, Levit $\&$ Stock 2002). The {\it in
vivo} experiment of Sourjik $\&$ Berg (2002) follows
the immediate changes in \cheyp\ dephosphorylation after step changes
in attractant concentration. These experiments show that the measured
decrease of the \chea\ activity with increasing attractant
concentration is functionally similar, but not identical, to the
decrease of the ligand-free receptor concentration. The experimental
curves of kinase activity as a function of  ligand concentration can
be fitted with Hill functions of the form

\begin{equation}
A(L)= A_0(1 -\Psi)=A_0\left(1-\dfrac{L^H}{K_A^H+L^H}\right),
\label{Hill}
\end{equation}
where $A$ represents the measured activity of the network, $A_0$ is
the maximal activity in the absence of ligand, $\Psi$ is the fraction of activity
suppressed by ligand binding, $L$ is the ligand
concentration, and $K_A$ is the ligand concentration that produces
half-maximal activity.  If we assume that there is no
interaction between receptors, the fraction bound with ligand is
\begin{equation}
\label{occupancy}
\theta_b =\frac{L}{K_D+L} = 1 -  \theta_f,
\end{equation}
where $K_D$ is the inverse of the affinity for ligand and $\theta_f$
is the fraction of receptors free of ligand.  If there are only two
possible states of the receptor complex, free and ligand-bound, and
only the former lead to autophosphorylation of CheA and a measurable
activity, then the activity would have the functional form 
$A(L)=A_0\theta_f$.  However, the experimental observations indicate a
more complex relationship, in that $K_A$ can be either larger or
smaller than $K_D$ and the Hill coefficient $H$ can be between $1$ and
$3$ (see Li $\&$ Weis 2000, Bornhorst $\&$ Falke 2001,
Sourjik $\&$ Berg 2002, Levit $\&$ Stock 2002).

If the activity is given in the form at (\ref{Hill}), we can compute
the relative change in activity $A(L)$ and the relative change in
receptor occupancy for a small change in ligand concentration.  Then
the gain, which we define as the ratio of relative changes, is given by
\begin{equation}
\label{gain}
g \equiv   \dfrac{d \ln A/dL}{d \ln \theta_b/dL} =-
\frac{H}{K_D}\frac{L^H (L+K_D)}{L^H+K_A^H} = -H\dfrac{\Psi}{\theta_f},
\end{equation}
and this is monotone increasing with $L$. Thus high amplification is
always possible for a sufficiently large ligand concentration, 
e.g. $L\gg K_A, K_D$ , and this conclusion holds even if the
ligand occupancy has a more complicated dependence on $L$, as long as
it approaches one for large $L$. The explanation of the high
amplification is clear from (\ref{gain}): at high ligand
concentrations the fraction of the activity suppressed $\psi$ approaches one,
whereas the fraction of receptors free of ligand $\theta_f$ approaches zero.
Thus the existence of high gain near saturation follows from the
functional form of the input-output relation of the upstream signal
transduction network, and even the simplest assumption of output
proportional to $\theta_f$ leads to high amplification for $L$ large
compared to $K_D$. Accordingly, the objective of a model should be to
predict the maximal activity $A_0$, the apparent dissociation constant
$K_A$ and the Hill coefficient $H$.

It is found experimentally that $A_0$, $K_A$ and $H$ depend on the
methylation state of the receptors and the presence or absence of the
methyltransferase CheR and the methylesterase CheB.  $A_0$
increases with the methylation level of the receptors and varies
approximately 30-fold (Li $\&$ Weis 2000, Bornhorst $\&$ Falke 2001, 
Sourjik $\&$ Berg 2002). $K_A$ also
increases with methylation state, and varies over two orders of
magnitude (Li $\&$ Weis 2000, Bornhorst $\&$ Falke 2001,
Sourjik $\&$ Berg 2002, Levit $\&$ Stock 2002),
which implies that the simplest assumption that ligand-free receptors
determine the output is not valid (Levit $\&$ Stock 2002). The Hill
coefficients of the output curves obtained in different experiments
vary between $1$ and $3$, and depend very weakly on the methylation
level. {\it In vivo} experiments also suggest that CheR and CheB have
a direct effect on the network output, in addition to determining the
methylation state of the receptors, because the CheR and CheB single
mutants show a qualitatively different response than CheRCheB mutants
with fixed methylation states (Sourjik $\&$ Berg 2002).

Our objective here is to propose a mechanism, based on receptor
clustering to form active teams, that can reproduce the
methylation-induced variability in the network output.  There are
several recent indications that the receptor-CheW-CheA complexes are
not static and do not have a one-to-one stoichiometry, as assumed
previously. Instead, an oligomer of multiple receptor dimers,
including different types of receptors, forms the core of an active
signaling complex (Ames et al. 2002, Francis et al. 2002). Since
chemotaxis receptors tend to be clustered at one end of a bacterium
(Maddock $\&$ Shapiro 1993), we assume that individual homodimers exist in
a dynamic equilibrium among singles, teams of two (twofolds), and
teams of three (threefolds), and that the distribution among st these
states depends on the ligand concentration.  Our central hypothesis is
that only threefolds can form complexes with CheW and CheA and
activate the autophosphorylation of CheA. Because the experimental
results we set out to explain all focus on the initial changes in
kinase activity, we do not consider the slower methyl-transfer
reactions. Since phosphotransfer from CheA to CheY is faster than the
autophosphorylation of CheA, the concentration of phospho-CheY is
proportional to the concentration of phospho-CheA, and the output of
the network is taken to be proportional to the concentration of
ligand-free threefolds in the model.

\noindent{\bf The Model}

\begin{figure}[htb]
\centerline{\psfig{file=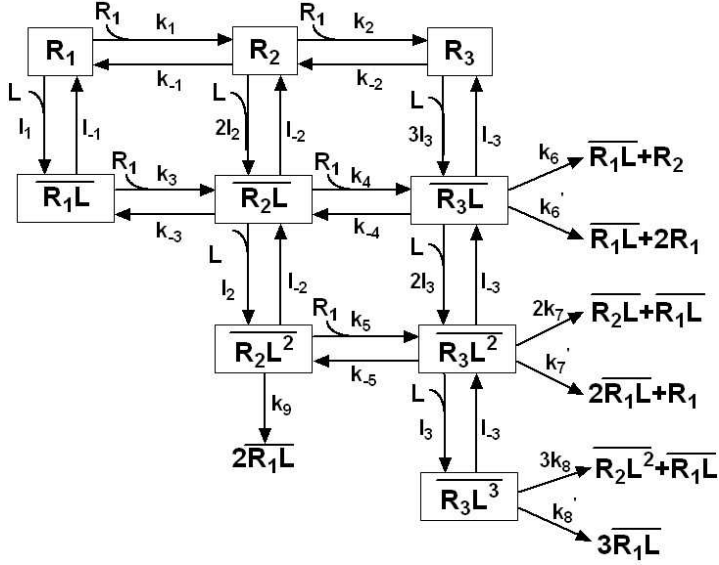,width=.55\textwidth}} 
\caption{The detailed reaction network for team formation and
ligand binding when there is only one type of receptor. Individual receptor dimers ($R_1$) can 
associate to form twofolds ($R_2$) and threefolds ($R_3$). Ligand binding to receptor teams leads to
the dissociation of the team. Only ligand-free threefolds can initiate kinase activity.}
\label{purescheme}
\end{figure}

The basic units of the model are receptor dimers, and we first
restrict attention to the inter-dimer association/dissociation and the
ligand-binding and release reactions for a single receptor type.
Homodimers are denoted by $ R_{1}, $ twofolds by $ R_{2} $, and
threefolds by $ R_{3}$ (see Figure
\ref{purescheme}). Receptor teams can have as many ligand-bound states
as there are receptor dimers in the team. For example, the
$\overline{R_2L} $ state contains a single ligand-bound receptor,
while $\overline{R_2L^2}$ has two ligand-bound receptors, one on each
dimer. We do not consider the state in which two ligand molecules are
bound to a homodimer because this is energetically unfavorable. We
assume that the ligand binding affinity of ligand-free homodimers in a
team is the same regardless of the binding state of other homodimers
in the same team\footnote{The proportionality factors $2$ and $3$ in
the ligand-binding reactions arise from combinatorial effects.}. We
allow for the possibility that dimers in receptor teams do not have
the same affinity for ligand as single receptor dimers ({\it i.e.}
$l_2$ and $l_3$ can be different than $l_1$). We assume that both
ligand-free and ligand-bound receptor dimers can associate to form
teams, possibly with different rates {\it i.e.} $k_1$ and $k_3$ can be either 
equal or different.

The main assumption of our model is that
ligand binding destabilizes receptor teams and consequently they break
into smaller units. We allow for every combination of resulting 
components, but assume that receptors will not release their
ligands in the process (see Figure \ref{purescheme}). In our model only the ligand-free
threefolds lead to CheA activation. As a result, kinase activity is proportional to the
concentration of $ R_{3} $, and its predicted dependence on ligand
concentration can be compared with the experimental results on kinase
activity.

\begin{figure}[htb] 
\centerline{
\psfig{file=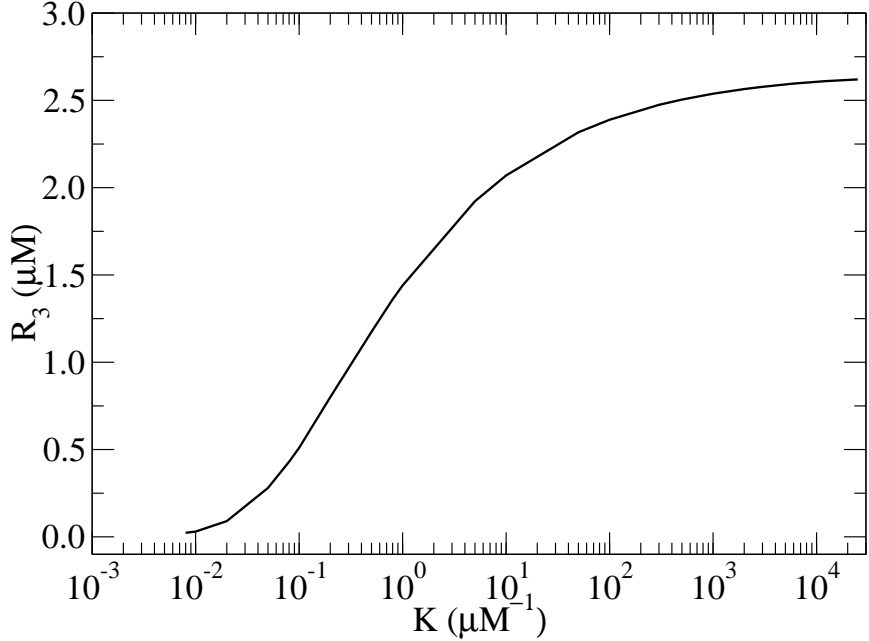,width=.6\textwidth,angle=-90} 
}
\caption{The dependence of the active team concentration on
the team association/dissociation characteristics as cumulated into
the parameter $K$. In the following we model different methylation
levels by choosing $K$ values such that the ratio of their
corresponding activities is  close to the experimental results of
Bornhorst and Falke (2001).}  
\label{methyl}
\end{figure}

The kinetic equations for the ligand-free states in Figure
\ref{purescheme} are as follows; equations for the remaining states
can be derived assuming mass-action kinetics.
\begin{align}
\label{fullmodel_eq}
\frac{dR_{1}}{dt}&=-2k_{1}R^{2}_{1}+2k_{-1}R_{2}-k_{2}R_{1}R_{2}+k_{-2}R_{3} 
-l_{1}R_{1}L+l_{-1}\overline{R_{1}L}+k_{-3}\overline{R_{2}L}+
k_{-4}\overline{R_{3}L}\nonumber \\
&-k_{4}R_{1}\overline{R_{2}L}+2k_{6}'\overline{R_{3}L}+
k_{-5}\overline{R_{3}L^{2}}-k_{5}R_{1}\overline{R_{2}L^{2}}+
k_{7}'\overline{R_{3}L^{2}}-k_{3}R_{1}\overline{R_{1}L}\\
\frac{dR_{2}}{dt}&=k_{1}R^{2}_{1}-k_{-1}R_{2}-k_{2}R_{1}R_{2}+
k_{-2}R_{3}-2l_{2}R_{2}L+l_{-2}\overline{R_{2}L}+
k_{6}\overline{R_{3}L}\nonumber\\ 
\frac{dR_{3}}{dt}&=k_{2}R_{1}R_{2}-k_{-2}R_{3}-3l_{3}R_{3}L+
l_{-3}\overline{R_{3}L}\nonumber 
\end{align}
If we define  the
equilibrium constant $K_{1}\equiv k_{1}/k_{-1}$ 
($K_{2}\equiv k_{2}/k_{-2}$), for the formation of twofolds 
(resp., threefolds); then at $ L=0,\, R_{3} $ satisfies the equation
\begin{equation}
3R_{3}+2\frac{K_{1}^{1/3}}{K_{2}^{2/3}} R^{2/3}_{3}+\frac{1}{(K_{1}
\cdot K_{2})^{1/3}}R^{1/3}_{3}-R_{T}=0. 
\label{eq_free}
\end{equation}
where $R_T$ is the total receptor concentration fixed at
$8 \mu M$. This equation has a unique positive root that tends to zero
as $K_{1}$ and/or $K_{2}$ tend to zero, approaches its maximum
$R_{T}/3$ as $K_{1}$ and/or $K_{2}$ tend to infinity, and increases
monotonically between these limits along any ray in the $K_1 - K_2$
plane.  Since the rates of team association/dissociation are not
known, we assume that $K_{1}=K_{2} \equiv K$ and choose
the individual rates
$ k_{1},k_{2},k_{-1},k_{-2} $ accordingly. Then  $R_3$ is completely
determined by $K$ for fixed $R_T$, and varies with $K$ as shown in      
Figure \ref{methyl}. 

There is a close parallel between Figure \ref{methyl} and the
experimental observations regarding the dependence of the ligand-free
output, $A_0$, on the methylation state of the receptors. The output
of the model, $R_3$, increases with the team formation constant $K$ in
a very similar way to the increase of $A_0$ with methylation. Notice
that the nonlinear increase of $R_3$ can explain both the observation
that the lowest methylation state's activity is only a fraction of the
activity of the wild type (Sourjik $\&$ Berg 2002) and the result that
the activity of the higher methylation states is very close (Levit
$\&$ Stock 2002). Based on this parallel, we identify different
methylation states with different choices for the parameter $K$, and
we choose these values such that the ligand-free activities
corresponding to these values have approximately the same proportions
as the experimental measurements of Bornhorst and Falke (Bornhorst
$\&$ Falke 2001). Thus we identify the unmethylated (EEEE) state with
$ K(0) = 10^{-2}\, \mu M^{-1}$ and the totally methylated (QQQQ) state
with the saturation limit $K(4)= 10^{3}\, \mu M^{-1} $. Bornhorst
and Falke constructed $16$ engineered states corresponding to all the
possible combinations of glutamate (E) and glutamine (Q)
residues. Based on their results, and similarly to other models (Spiro
et al. 1997, Barkai $\&$ Leibler 1997, Morton-Firth et al. 1999) we
assume that the total methylation level is the crucial characteristic
of a given state, and not the exact residues that are methylated.  We
choose the team formation constants for partially methylated states as
follows: methylation level one (e.g. EEEQ) $K(1)= 10^{-1}\, \mu M^{-1}$,
methylation level two (e.g. QEQE) $ K(2) = 1 \,\mu M^{-1}, $ and
methylation level three (e.g. QQQE) $K(3) \equiv 10 \,\mu M^{-1} $.

In order to compare our results with the experimental results in
Bornhorst $\&$ Falke (2001) and Sourjik $\&$ Berg (2002), we assume that
the receptor is Tar and that the ligand is methyl-aspartate. It is
known that the affinity of Tar to methyl-aspartate is about ten times
less than to aspartate; therefore, we assume that the ligand release
rate of a single receptor dimer is ten times larger than the release
rate of aspartate, which is $70 s^{-1}$, while the ligand binding rate is the same
as the binding rate to aspartate, which is $70 \mu M^{-1} s^{-1}$
(Spiro et al. 1997). Correspondingly, we assume 
$l_1=70\mu M^{-1}s^{-1}$, $l_{-1}=700 s^{-1}$. We assume that the ligand release 
rates of receptor twofolds and threefolds are the
same as the release rate of a single receptor dimer, {\it i.e.} $l_{-2}=l_{-3}=700 s^{-1}$. 

We consider that the
association rate of $R_1$ with $R_2$ is slightly smaller than the
association between two $R_1$s, and the dissociation rate of $R_3$ is
proportionally smaller than the dissociation rate of $R_2$, such that
the ratios $K_1\equiv k_{1}/k_{-1}$ and $K_{2}\equiv k_{2}/k_{-2}$ are
equal. Thus we choose the rates to be $k_{-1}=0.1s^{-1}$, 
$k_{-2}=0.05s^{-1}$, and we vary $ k_{1}$ and $k_{2} $ according to 
$ k_{1}=10^{-2+d}\,\mu M^{-1}s^{-1},k_{2}=5\cdot 10^{-3+d}\,\mu M^{-1}s^{-1}$ 
for $d=0,1,2,4 $, such that $K$ 
corresponds to the different methylation levels described above.

As there is no information about the relative rates with which free or ligand bound
receptor dimers associate/dissociate, we assume that the ligand binding state does not
influence team formation and therefore $k_{3}=k_{1}$, $k_{-3}=k_{-1}$ 
and $k_{5}=k_{4}=k_{2}$, $k_{-5}=k_{-4}=k_{-2}$. For the breakdown of 
ligand-bound threefolds we consider that processes involving a single 
dissociation are equiprobable, {\it i.e.} $k_{6}=k_{7}=k_{8}=k_{9}=0.7s^{-1}$,
while the processes involving two dissociations are less likely, 
$k_6'=k_7'=k_8'=0.07 s^{-1}$.

\begin{figure}[htb]
\centerline{\psfig{file=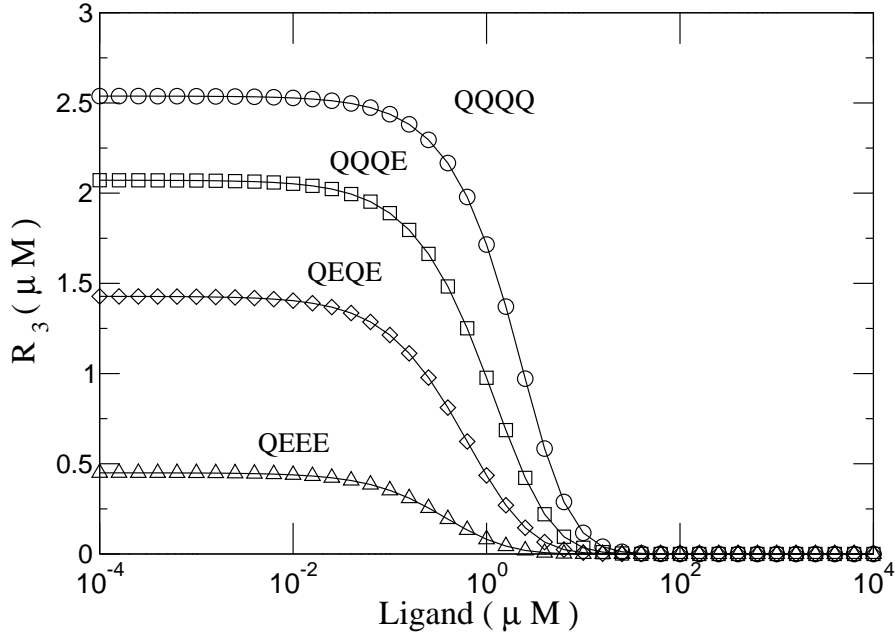, width=.6\textwidth,angle=-90} 
}
\caption{ Concentration of ligand-free threefolds, $R_3$, as a
function of external ligand concentration for four different
methylation levels, assuming that receptors in teams have the same
ligand binding affinity as isolated receptors. Circles, $K=1000 \,
\mu M^{-1}$ (QQQQ); squares, $K=10 \, \mu M^{-1}$ (QQQE); diamonds,
$K=1 \mu M^{-1}$ (QEQE); triangles, $K=0.1 \mu M^{-1}$ (QEEE). The
continuous lines represent fits of the form (\ref{Hill}). The
apparent dissociation constants and Hill coefficients are QQQQ --
 $ K_{A}=1.65 \, \mu M$, $ H=1.37 $; QQQE -- $ K_{A}=0.84 \, \mu M $, $ H=1.23
$; QEQE -- $ K_{A}=0.48 \, \mu M $, $ H=1.19 $; QEEE -- $ K_{A}=0.30 \,
\mu M $, $ H=1.23 $.} 
\label{fig4}
\end{figure}

First we assume that the ligand binding rate of dimers that are part
of receptor teams is the same as the ligand binding rate of separate
receptor dimers, {\em i.e.}, $l_2=l_3=l_1$. Figure \ref{fig4} shows the
steady-state value of $R_3$ as a function of the ligand concentration
for four different $ K $ values corresponding to four methylation
levels.  These curves are obtained by solving the entire system of
steady-state equations using the software package AUTO
(Doedel 1981).   All curves can be fit with Hill functions of the form
(\ref{Hill}), wherein $ A(L)\equiv R_{3}(L)$.

This figure captures many features of the experimentally-observed
decay in kinase activity for increasing ligand concentrations. The
curves are in qualitative agreement with those reported in Li $\&$
Weis (2000), Bornhorst $\&$ Falke (2001), Sourjik $\&$ Berg (2002),
and Levit$\&$ Stock (2002); the apparent $K_{A}$ increases consistently with
methylation level while the Hill coefficient does not. Thus our
scheme provides a possible explanation for the apparent dependence of
the receptor affinity on the receptor methylation level. We do
not change the true affinity for ligand, but varying the affinity of
receptor dimers for other receptor dimers leads to the differential
response of kinase activity to ligand. Unfortunately, the $K_{A}$
values predicted by the model are lower than the experimental results
obtained for CheRCheB mutants, and the range of their variation is
also much smaller. Note, however, that the wild type response measured
by Sourjik and Berg (2002) exhibits a small $K_A=1 \,\mu M$,
close to our result of $K_A = 0.48 \,\mu M$.

\begin{figure}[htb]
\centerline{
\psfig{file=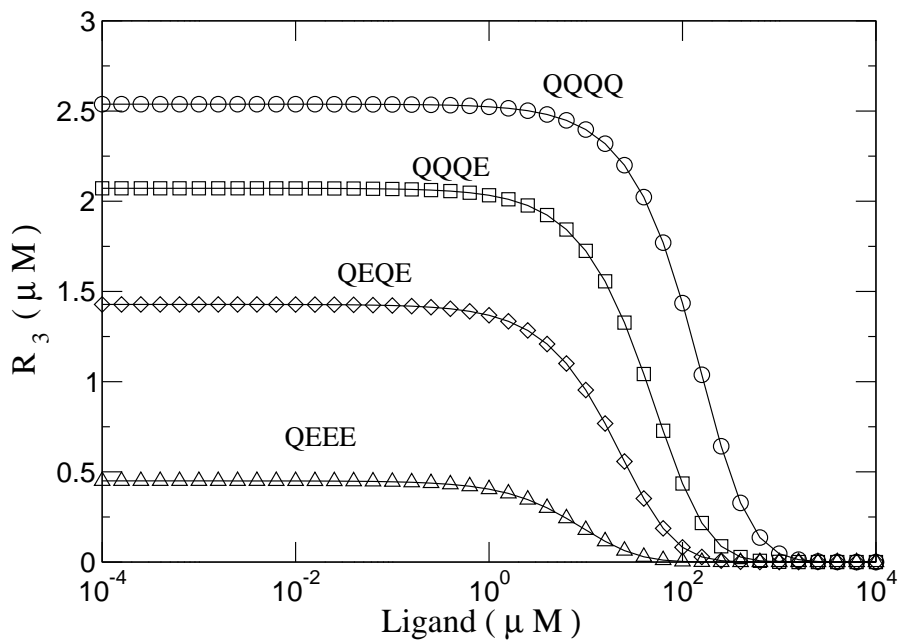, width=.6\textwidth,angle=-90} }
\caption{Concentration of ligand-free threefolds, $R_3$, as a
function of external ligand concentration for four different
methylation levels, assuming that the affinity of receptors in teams
is 1\% of that of isolated receptors.  The $ K_{A} $ values and Hill
coefficients are: QQQQ -- $ K_{A}=112.98\,\mu M$, $ H=1.38 $;
QQQE -- $ K_{A}=37.20 \, \mu M $,  $ H=1.32 $; QEQE -- $ K_{A}=16.57 \,
\mu M$,  $ H=1.30 $; QEEE -- $ K_{A}=6.68 \, \mu M $, $ H=1.28 $. }
\label{fig5}
\end{figure}

It is easily seen that the larger $K_A$ values observed in experiments
on CheRCheB mutants could be explained by assuming that receptor
teams have a lower affinity for ligand than individual receptor
dimers. The cause of this lowered affinity could be the 
close proximity of receptors in teams. To illustrate this case we consider 
$l_{2}=l_{3}=l_1/100$, while keeping all other parameters at their
previous values. The resulting activity curves as a function of the
ligand concentration are shown in Figure \ref{fig5}.

These curves agree well with the results given in Figure 2(c) of
Bornhorst and Falke (2001). The  Hill coefficients
in (Bornhorst and Falke 2001) range from $1.1$ to $2.2$, and our values
are in this range; the range of $ K_{A} $ values are from 
$15 \, \mu M$ (QEEE) to $97 \,\mu M$ (QQQQ), ours are around this
range, too. Moreover, the apparent $ K_{A} $ increases dramatically
with methylation level, while the value of $ H $ is about the same for
each methylation level; which is consistent with all the experimental
results (Li $\&$ Weis 2000, Bornhorst $\&$ Falke 2001, Sourjik $\&$ Berg 2002, 
Levit $\&$ Stock 2002). 

Comparing Figure \ref{fig4} and Figure \ref{fig5}, we can notice that
$K_{A}$ depends strongly and inversely on the ligand-binding rates of the
receptor teams. Consequently, a possible effect of CheR and/or CheB that would
modify the ligand-binding affinity of receptor teams would explain the
methylation-independent variation of $K_A$.

In order to better understand the dynamics of the reaction network, we
consider the changes in the 
concentrations of  different states on  Figure
\ref{purescheme} under changes in parameters and the ligand
concentration. Because the  total amount of receptor is fixed, changes in
ligand  propagate through the network until a
new steady state is reached. In Table \ref{states} we compare the
concentrations of the ligand-free and completely occupied states for two
external ligand concentrations, two different methylation levels, and two types
of team ligand binding behaviors.

As we saw earlier, at  $L=0$ the proportions of
concentrations in these states depends strongly on the methylation state; in the
highly methylated case the vast majority of receptors are in threefolds,
while in the wild type methylation case the states are equilibrated. The
addition of $L= 100 \, \mu M$ induces dramatic changes in the state occupancies.
These changes depend both on the methylation level and the ligand affinity of
receptor teams. For $l_1=l_2=l_3$, ligand-free teams all but disappear, in both
methylation states. The majority of the receptors are now in the ligand-occupied
states. In the high methylation case the totally ligand-bound teams are most
abundant, while in the wild type methylation case almost every receptor is
in the ligand-bound single dimer state. For $l_2=l_3=l_1/100$, the effect of
ligand is much weaker, and it depends crucially on the methylation state. At high
methylation, the ligand-bound states are sparsely populated, and the majority of
receptors is in the ligand-free teams, $R_2$ and $R_3$. Notice that the
concentration of $R_2$ almost quadruples, and the moderate value of $\overline{R_1L}$ 
is the only indication of the presence of ligand. In the wild
type methylation state the most occupied states are the ligand-bound isolated
receptors and, still, ligand-free intermediary teams. We can thus conclude that
in the $l_1=l_2=l_3$ case the most important response to ligand is a
vertical flow from ligand-free to ligand-bound states, while in the
$l_2=l_3=l_1/100$ case the most important flow is a horizontal one from
receptor teams to individual receptor dimers.

The complexity of the state space induced by team formation also
raises the question whether the fraction of ligand-bound states is
dependent on the methylation level, and how it compares to the ligand
affinity of individual receptor dimers. We have calculated the
dependence of ligand occupancy,
$(\overline{R_{1}L}+2\overline{R_{2}L}+3 $
$ \overline{R_{3}L}+2\overline{R_{2}L^{2}}+3\overline{R_{3}L^{2}}+3\overline{R_{3}^{}L^{3}})/R_{T}$
on $L$, and find that it depends on the assumptions about team
affinity and methylation state. For example, we find that the apparent
dissociation constant corresponding to wild type methylation in the
$l_{1}=l_{2}=l_{3}$ case is slightly smaller than the $K_D=10 \, \mu M$ 
of individual receptor dimers, while in the $ l_{2}=l_{3}=l_1/100$
case the apparent dissociation constant is higher than $K_D$.

\begin{table}[htb]
\begin{tabular}{|c|c|c|c|c|c|c|c|c|}
\hline 
&
K&
L&
$R_{1}$&
$R_{2}$&
$R_{3}$&
$\overline{R_{1}L}$&
$\overline{R_{2}L^{2}}$&
$\overline{R_{3}L^{3}}$\\
\hline
$l_1 = l_2 =$&
$10^{3}$&
0&
0.0136&
0.1861&
2.5381&
0&
0&
0\\ 
\cline{3-9}
 & 
$\mu M^{-1}$&
100&
0.0331&
0.0079&
2.0933e-4&
0.3366&
1.5849&
1.2445\\
\cline{2-9} 
$l_3 = 70\mu M^{-1}$&
1&
0&
1.1337&
1.2909&
1.4282&
0&
0&
0\\ 
\cline{3-9} 
& 
$\mu M^{-1}$&
100&
0.6094&
0.0029&
1.3973e-6&
6.0943&
0.5740&
0.0083\\
\hline 
$l_2 = l_3 =$&
$10^{3}$&
0&
0.0136&
0.1861&
2.5381&
0&
0&
0\\ 
\cline{3-9}
& 
$\mu M^{-1}$&
100&
0.0115&
0.8169&
1.4353&
0.1156&
0.0164&
0.0086\\
\cline{2-9}
$l_1/100 = 0.7\mu M^{-1}$&
1&
0&
1.1337&
1.2909&
1.4282&
0&
0&
0\\ 
\cline{3-9}
& 
$\mu M^{-1}$&
100&
0.4037&
1.3217&
0.0815&
4.0374&
0.0264&
4.8547e-4\\
\hline   
\end{tabular}
\caption{ The effect of changes in parameters and ligand levels on the
distribution of states in the network. $K = 1 \mu M$ corresponds to
QEQE (wild type), whereas $K = 10^{3} \mu M$ corresponds to QQQQ. All concentrations are
measured in  $\mu M$.} 
\label{states}
\end{table}

Finally, we study how the different rates $ k_{i} $ and $ l_{i} $
change the activity curves of the model.
\begin{enumerate}

\item{In our choice of parameters we assumed that the ligand  binding state does not influence 
the receptor association rate. To 
explore the effects of different rates, we test the extreme situation when one sets the rates 
$k_{3} \mbox { and } k_{-3} $, $ k_{4}\mbox { and } k_{-4} $ or 
$k_{5}\mbox { and } k_{-5}$ to zero with the other parameters and conditions fixed. We find that 
for each case individual concentrations were
altered but there was little effect on $ R_{3}$. This suggests that the equality of these rates is
not a strict condition for the success of the model, and the association between ligand free receptors
has the dominant effect on $R_3$}

\item{In our model we assumed that the ligand-free threefolds constitute the kinase-activating
state. To test whether our conclusions are generally valid for teams comprised of different numbers
of receptor dimers, we assume that the association of twofolds with individual dimers is prohibited,
and ligand-free twofolds are the kinase-activating state. In other words, we set $k_2=k_4=k_5=0$. 
We find that $R_2(L=0)$ follows a curve very similar to Figure \ref{methyl}, with the only difference
that the saturation value for high $K$ is $R_0/2$ instead of $R_0/3$. Selecting the same $K$
values for the different methylation levels as before we obtain that $K_A$ varies between
 $2.22 \mu\,M$ (QEEE)
and $3.44 \mu\,M$ (QQQQ) for $l_1=l_2$ and in the range $25.3 \mu\,M$  (QEEE) - $314 \mu\,M$ (QQQQ) 
for $l_2=l_1/100$. The closeness of these results to our original 
model suggests that the number of steps involved in kinase-activating team formation does not 
have a crucial role.}

\item{The effect of the single receptor ligand-binding rate $l_{1}$ on the
activity curve is not as strong as the ligand-binding rates for receptor teams.
When we assume $l_{1}=l_{2}=l_{3}=0.7 \,\mu M^{-1}s^{-1}$ rather than 
$l_{1}=70 \,\mu M^{-1}s^{-1}$, $l_{2}=l_{3}=l_1/100$ as in 
Figure \ref{fig5}, $K_A$ increases considerably, but not as much as the 
change between Figure \ref{fig4} and Figure \ref{fig5}. This suggests that an overall
less-than-expected affinity to methyl-aspartate might be at the root of the large
observed $K_A$ values.} 
\end{enumerate}      

In conclusion, our results show that a model based on active threefolds of a pure
receptor can explain the {\it in vitro} activity curves
(Li $\&$ Weis 2000, Bornhorst $\&$ Falke 2001, Levit $\&$ Stock 2002). The assumptions of
methylation dependent dynamic team formation and ligand-induced breakdown lead to differential kinase
activity curves without invoking methylation-induced changes in ligand affinity. 
We turn next to the experimental observations on mixed receptor types (Sourjik and Berg 2002).

\medskip\noindent{\bf Mixed Receptor Types} 

Bacteria have several types of receptors, and it is possible that different types of 
receptor interact in order
to be able to respond optimally to diverse environmental stimuli. Indeed, the
experiments of Sourjik and Berg (2002) suggest that under
certain conditions both the Tar
and Tsr receptors respond to methyl-aspartate. According to these
experiments, 
CheRCheB mutants with fixed methylation levels have two apparent dissociation
constants corresponding to the Tar and Tsr receptors, respectively, and can be
fit by Hill functions of the form
\begin{equation}
\dfrac{A(L)}{A_0} = 1-\beta\frac{L^{H_T}}{L^{H_T}+K_T^{H_T}}-(1-\beta)
\frac{L^{H_S}}{L^{H_S}+K_S^{H_S}}.
\label{multi_Hill}
\end{equation} 
The simplest model suggested by these results is based on the
assumption that the output of the composite system is the sum of two
individual outputs similar to (\ref{Hill}), one for each of the  two
pure receptor populations. For this model the total output is 
\begin{eqnarray}
\dfrac{A(L)}{A_0} = 1-\frac{A_{T0}}{A_0}\frac{L^{H_T}}{L^{H_T}+K_T^{H_T}}-
\frac{A_{S0}}{A_0}\frac{L^{H_S}}{L^{H_S}+K_S^{H_S}},
\label{mixed}
\end{eqnarray}
where $A_0=A_{T0}+A_{S0}$.
If the Tsr methylation level is constant, as it appears to be
experimentally (Sourjik $\&$ Berg 2002), Equation \ref{mixed} predicts that an
increase in the Tar methylation level leads to an increase in $\beta\equiv A_{T0}/(A_{T0}+A_{S0})$.
However, the experiments indicate that for the CheRCheB mutants, $\beta$ decreases
from $0.65$ to $0.27$ as the methylation state of the Tar receptor
changes from EEEE to QQEQ. Moreover, $\beta$ appears to
be $1$ for CheR mutants  that are in the lowest methylation
state and it is $0$ for the CheB mutants that are in the highest
methylation state (Sourjik $\&$ Berg 2002).  Therefore the experimental
results cannot be explained if it is assumed that receptors act
independently, which strongly suggests that there are interactions
between different receptor types, in addition to the interactions
within pure types. This leads to the possible formation of mixed
teams, and in the following we determine whether the model for team
formation of pure types, extended to two types of receptor, is able to
generate response curves similar to those in (Sourjik $\&$ Berg 2002).

We denote the two types of receptors by R and P, and assume that they
have different affinities for ligand. We also assume that two receptor
dimers can associate to form pure or mixed receptor teams, and in the
general case the association/dissociation constants of two R (resp, P or R
and P) receptor dimers, $K\equiv k_1/k_{-1}$, (resp., $H\equiv h_1/h_{-1}$ and
$M\equiv m_1/m_{-1}$), are different. 

\begin{figure}[htb]
\centerline{
\psfig{file=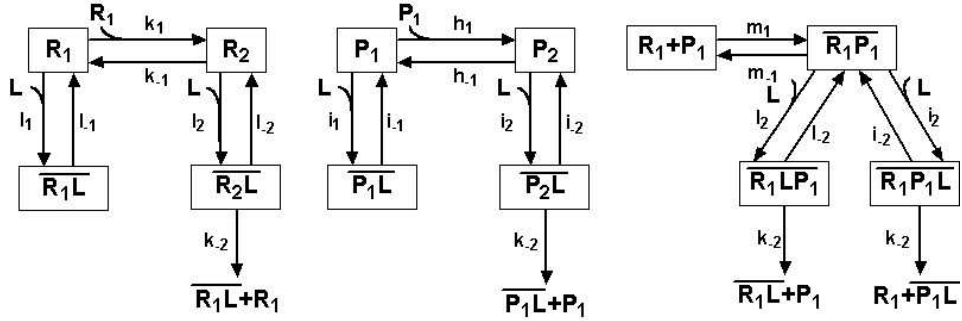, width=.25\textwidth,angle=-90}
}
\caption{The kinetic scheme for formation of pure and
mixed receptor twofolds.  Both receptor types  bind ligand, but with
different affinities, and binding induces team dissociation.
}
\label{mix_scheme}
\end{figure}

The full scheme of all mixed twofold and threefold states and the
transitions between them includes $29$ states instead of the $9$ in
Figure \ref{purescheme}, and involves $38$ unknown rates for the
receptor association/dissociation reactions alone. It is not
worthwhile to tackle this level of computational complexity in the
absence of any experimental information, and our previous analysis
suggested that a reduced scheme with receptor twofolds as the
ligand-activating state leads to similar results as the original
scheme. Thus, to reduce the complexity of the analysis of mixed team
formation, we only consider receptor twofold formation, and we do not
consider teams with more than one ligand-bound receptor. Consequently,
the reaction scheme contains two types of receptor dimers, $R_1$ and
$P_1$, the ligand-bound states of these dimers, three different
twofolds, $R_2$, $\overline{R_1P_1}$ and $P_2$, and six ligand-bound
states of these teams, $\overline{R_1L}$, $\overline{R_2L}$,
$\overline{R_1LP_1}$, $\overline{R_1P_1L}$, $\overline{P_1L}$ and
$\overline{P_2L}$. As in the case of pure receptors, we allow for
the possibility that teams have lower ligand affinity than
homodimers.  We assume that binding of ligand to either dimer in a
twofold induces the dissociation of the twofold, and that the rate of
dissociation is the same for all teams.  Figure \ref{mix_scheme} shows
the entire kinetic scheme and the associated rates.

This scheme leads to the following steady state equations.
\begin{equation}
\label{eq_R}
\frac{2k_1(1+a_2L)}{k_{-1}+k_{-2}a_2L}R_1^2 +\frac{2m_1[1+(a_2+b_2)L]}
{m_{-1}+k_{-2}(a_2+b_2)L}R_1 P_1+(1+a_1L)R_1=R^t
\end{equation} 
\begin{equation}
\frac{2h_1(1+b_2L)}{h_{-1}+k_{-2}b_2L}P_1^2 + \frac{2m_1[1+(a_2+b_2)L]}
{m_{-1}+k_{-2}(a_2+b_2)L}R_1P_1+(1+b_1L)P_1=P^t
\label{eq_P}
\end{equation}

where $a_1=l_1/l_{-1}$ is the affinity of a single R dimer for ligand,
$b_1=i_1/i_{-1}$ is the affinity of a single P dimer,
$a_2=l_2/l_{-2}$ is the affinity of an R dimer that is part of a (pure or
mixed) twofold, and $b_2=i_2/i_{-2}$ is the affinity of a P dimer that is part of
a twofold.

We assume that the kinase-activating output of this system is the concentration
$R_2+\overline{R_1P_1}+P_2$ of free twofolds. Expressing each of these terms as a
function of single receptor concentration we obtain

\begin{equation}
A(L)=\dfrac{k_1R_1^2}{k_{-1} + k_{-2}a_2L} +
\dfrac{2m_1R_1{P_1}}{m_{-1} + k_{-2}(a_2+b_2)L} 
+ \dfrac{h_1P_1^2}{h_{-1}+h_{-2}k_2L}
\end{equation} 

The main differences between this output and the simple assumption of
non-interacting receptors are the existence of the second term depending on
$R_1P_1$, and the fact that the steady state concentrations of $R_1$ and $P_1$ are
coupled.

Equations (\ref{eq_R}) and (\ref{eq_P}) can be solved numerically to
obtain the output of the network as a function of the ligand
concentration.  To account for the results reported in
(Sourjik $\&$ Berg 2002), we assume that the P receptor corresponds to
Tsr and its affinity for methyl-aspartate is $10^3$ times lower than
the affinity of the R receptor (Tar). The other rates are chosen to
correspond with the rates used in the pure population. Thus we set 
$a_1=0.1\,\mu M^{-1}$, $b_1=10^{-4}\,\mu M^{-1}$,
$k_{-1}=h_{-1}=m_{-1}=0.1\, s^{-1}$, $k_{-2}=70\,s^{-1}$, 
and we allow $k_1$ to vary between $10^{-3} \mu M^{-1} s^{-1}$ (EEEE)
and $10^2 \,\mu M^{-1}  s^{-1}$ (QQQQ).

\begin{figure}[htbp]
\centerline{
\psfig{file=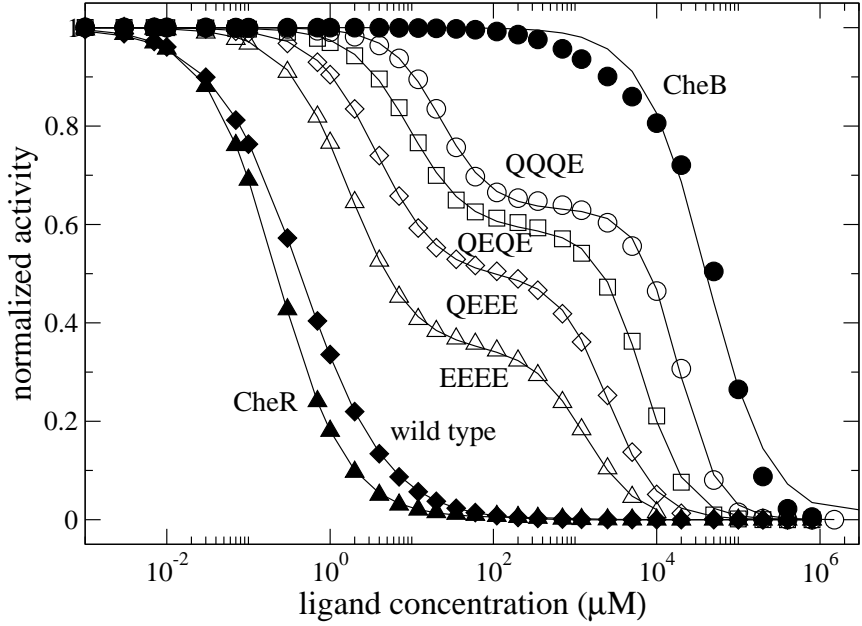, width=.6\textwidth,angle=-90}
}
\caption{ Activity of modeled Tar-Tsr mixtures 
as a function of ligand  
concentration. The Tsr receptors are assumed to be twice as abundant as
the Tar receptors, and their affinity to methyl-aspartate $10^3$
less. Open symbols stand for CheRCheB mutants, and have $a_2=a_1/100$
and $b_2=b_1/100$. Open circles, $K\equiv k_1/k_{-1}=10 \,\mu M^{-1}$ (QQQE),
$H\equiv h_1/h_{-1}=6\,\mu M^{-1}$, $M\equiv m_1/m_{-1}=10\,\mu M^{-1}$. Open squares, 
$K=1\,\mu M^{-1}$ (QEQE), $H=0.5\,\mu M^{-1}$, $M=1\,\mu M^{-1}$. Open diamonds, 
$K=0.1\,\mu M^{-1}$ (QEEE) $H=0.05\,\mu M^{-1}$, $M=0.1\,\mu M^{-1}$. Open triangles,
$K=10^{-2}\,\mu M^{-1}$ (EEEE), $H=0.005\,\mu M^{-1}$, $M=10^{-2}\,\mu M^{-1}$. The
curves can be fit by Hill functions like Equation (\ref{multi_Hill})
with two transition regions.  QQQE -- $\beta= 0.36$,
$K_T=22.5\,\mu M$, $H_T=1.38$, $K_S=18.2\, mM$, $H=1.76$. QEQE -- 
$\beta= 0.41$, $K_T=9.44\,\mu M$, $H_T=1.21$, $K_S=6.5\, mM$,
$H=1.57$. QEEE -- $\beta= 0.5$, $K_T=3.68\,\mu M$, $H_T=1.13$,
$K_S=2.4 \,mM$, $H=1.43$. EEEE --, $\beta= 0.65$,
$K_T=1.67\,\mu M$, $H_T=1.09$, $K_S=1.3 \,mM$, $H=1.35$.  Filled
diamonds (wild type), $K=M=1\,\mu M^{-1}$,
$H=10^{-4}\,\mu M^{-1}$,$a_2=a_1$, $b_2=b_1$.  Filled triangles (CheR
mutant), $K=M=10^{-2}\,\mu M^{-1}$, $H=10^{-4}\,\mu M^{-1}$, $a_2=a1/10$, $b_2=b_1/10$. Filled circles (CheB mutant),
$K=M=10^{6}\,\mu M^{-1}$, $H=100\,\mu M^{-1}$, $a_2=a_1/100$ and
$b_2=b_1/100$. These curves can be fit with single Hill functions like
Equation \ref{Hill}. The $K_A$ values and Hill coefficients are wild type -- 
$K_A=0.43 \,\mu M$, $H=0.82$; CheR mutant -- $K_A=0.22 \,\mu M$, $H=1.01$; 
CheB mutant -- $K_A=40\, mM$, $H=1.1$. The ratio of
amplitudes is wt:CheR:EEEE:QEEE:QEQE:QQQE:CheB=1:0.26:0.35:0.94:1.35:1.52:1.6.}
\label{fig_mix}
\end{figure}

To model the wild type activity curve, we assume that 
Tar is in its QEQE methylation state, and set $k_1=0.1\,\mu M^{-1} s^{-1}$. 
To capture the surprisingly fast decay of the wild type activity, we assume that
receptor teams have the same ligand binding affinity as single receptor dimers, 
{\em i.e.}, $a_2=a_1$ and $b_2=b_1$. We also  assume that the association rate of
Tsr into pure Tsr teams is lower than the association rate of Tar,
{\em i.e.}, $h_1=k_1/100$. These assumptions lead to  a Hill function of type 
\ref{Hill} with a low $K_A$, in good agreement
with the experimental results {Sourjik $\&$ Berg 2002} (see filled
diamonds in Figure \ref{fig_mix}). In the CheR mutant
both receptors are in their lowest methylation levels since they lack the methylating
enzyme but have the demetylating enzyme CheB. Again, we assume that the
association rate of Tsr is lower than that of Tar, {\em i.e.},
$k_1=m_1=10^{-3} \,\mu M^{-1} s^{-1}$ (EEEE) and $h_1=10^{-5}\, \mu
M^{-1} s^{-1}$.  The decrease in Tar methylation state induces the
decrease of both the ligand-free activity and the apparent
dissociation constant; however, the experiments indicate that the
$K_A$ of the CheR mutant is close to the $K_A$ of the wild type
curve. We are able to reproduce this result by assuming that teams
have a slightly lower ligand affinity than single receptors,
i.e. $a_2=a_1/10$ and $b_2=b_1/10$ (see filled triangles in Figure
\ref{fig_mix}).

Next we consider the CheRCheB mutants, and assume that $a_2=a_1/100$,
$b_2=b_1/100$.  We find that for the majority of choices for $k_1$,
$h_1$ and $m_1$ the output curves can be fit by generalized Hill
functions like Equation \ref{multi_Hill} with two fast-decaying
regions characterized by apparent dissociation constants that are
several orders of magnitude apart. We identify the lower dissociation
constant, $K_T$, with the Tar receptor, and the higher, $K_S$, with
the Tsr receptor.

The experimental results indicate that both the ligand-free output and
the apparent dissociation constants $K_T$ and $K_S$ increase with
increasing Tar methylation levels. Additionally, the parameter $\beta$, indicating
the relative weight of the Tar receptors in the output, decreases with
increasing Tar methylation levels. Our model results in a good
agreement with these conclusions if we assume that interaction between
the two receptor types leads to a moderate variability of the Tsr team
formation rate. The curves marked by open symbols on Figure
\ref{fig_mix} present our results for four different sets of
pure/mixed team formation rates. Different curves have Tar association
rates corresponding to different methylation states from EEEE to
QQQE. We assume that the Tsr-Tsr association rates are lower than the
Tar and Tar-Tsr association rates, and they also increase with Tar
methylation, but with a slower rate\footnote{Our studies indicate that
if we keep the Tsr team association/dissociation rates constant, the
high-ligand tails of the activity curves coincide, resulting in
normalized activity curves that have a reversed order compared to the
experimental curves, thus the only way to capture the systematic
upwards shift with methylation is by assuming a variation of the Tsr
methylation levels.}. We find that under these assumptions our results
are in excellent qualitative and good quantitative agreement with the
CheRCheB mutant results of Sourjik and Berg (2002).

In a CheB mutant the receptors are in the highest methylation state, since the action of
the methylating enzyme CheR is not balanced by CheB. We assume
that Tar has a very high association rate, both in pure and mixed teams, while
the Tsr-Tsr association rate is somewhat lower. As in the case of
CheRCheB mutants, we assume that $a_2=a_1/100$ and
$b_2=b_1/100$. We find that the output of 
such mixture has a single apparent
dissociation constant in the milimolar range (see filled circles in Figure
\ref{fig_mix}).

\noindent{\bf Discussion} 

We have shown that the high upstream sensitivity of the signal
transduction network is caused by the negative regulation between
ligand occupancy of the receptors and kinase activity. Since  kinase
activity decreases with increasing ligand occupancy, at sufficiently
high attractant concentrations the relative change in kinase activity
is much larger than the relative change in occupancy.  A related
general argument indicates that the sensitivity of the signal
transduction network, defined as the relative change in kinase
activity in response to a certain percentage change in ligand
concentration, depends only on the kinase activity suppressed by
ligand
\begin{equation}
S=-H\psi=-H\dfrac{L^H}{K_A^H+L^H},
\end{equation}
 and consequently 
approaches $-H$ at $L\gg K_A$. This implies that the marked differences between the sensitivity 
of the wild type and CheRCheB mutants found by Sourjik and Berg (2002)
are caused by the fact that for the ambient ligand concentrations studied the wild type receptors 
have reached the maximum sensitivity, while the CheRCheB mutants have not.

To illustrate this point we include a figure depicting the absolute value of the
sensitivity of two receptor populations
to a $10\%$ change in ligand concentration (similar to Figure 3b in Sourjik and Berg 2002). 
The first population's kinase response is described by a Hill function like Equation 
\ref{Hill} with the parameters 
$K_A=1\mu M$, $H=1$ while the second's kinase response follows Equation \ref{multi_Hill} 
with $K_{T}=150 \mu M$, $K_{S} = 100 mM$ and $H_T=H_S=1$.
In the ambient ligand concentration range $10 \mu M<L<10^4 \mu M$ the sensitivity of the first 
population is constant since $L>K_A$, while the second population, having $L<K_S$, has a varying 
and much smaller sensitivity. This behavior 
is in excellent qualitative agreement with the experimental observations of Sourjik and Berg (2002).

\begin{figure}[htb]
\centerline{\psfig{figure=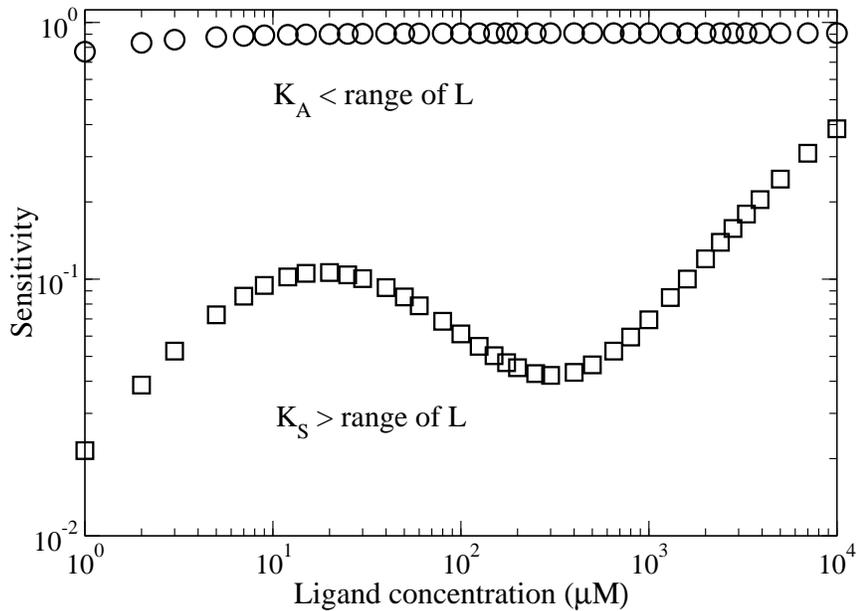,width=10cm,angle=-90}}
\caption{Sensitivity of two receptor populations to a $10\%$ increase in ligand concentration. 
The sensitivity
is defined as the ratio of the relative change in kinase activity and the relative change in 
ligand concentration.}
\end{figure}

We have demonstrated that a model based on the assumption that homodimers of
a receptor must aggregate into teams of three in order to activate the
autophosphorylation of CheA can adequately explain the observed
dependence of the kinase activity on the ligand concentration for a
pure receptor. Our model is in qualitative agreement with 
the experimental results, and shows that methylation-dependent kinase activity does not
necessarily imply methylation-dependent ligand affinity. We also showed that the concentration
corresponding to half-maximal kinase activity need not coincide with the apparent 
ligand dissociation constant of the receptor population, nor does the latter coincide with the
dissociation constant of an isolated homodimer.

Our model assumes that receptor populations possess a dynamic balance between homodimer,
twofold and threefold states, as opposed to an ordered threefold structure. This 
prediction, along with our assumptions for team formation and dissociation rates could 
be tested experimentally in {\it in vitro} receptor preparations. Furthermore, in order to
quantitatively reproduce the experimental results on CheRCheB mutants
(Bornhorst $\&$ Falke 2001) within the framework of
the detailed model it is necessary that twofolds and threefolds of receptors
have a lower affinity for ligand than an isolated homodimer. This theoretical
prediction could be verified experimentally by testing the affinity of homogeneous receptor 
preparations (i.e. only dimers or only teams).

When there are multiple receptor types, the experimentally-determined
activity curves display complex dependence on the ligand
concentration, but they can be satisfactorily reproduced by our
model. One consistent assumption that was needed is that the
association rate of Tsr teams is lower than the association rate of
Tar and Tar-Tsr teams. This assumption was vital in reproducing the
wild type, CheR and CheB mutant curves, and it suggests the existence
of receptor specificity in team formation capabilities. This feature
could be caused by receptor-specific methyl-accepting activities that
were confirmed experimentally (Barnakov et al. 1998).

Our results also confirm earlier suggestions that changes in methylation state (or
association/dissociation rates) alone cannot explain the 
qualitative difference between the wild type and CheRCheB mutant
activity curves. We were able to reproduce the shift by assuming that
in CheR or CheB mutants receptor teams have lower affinity for ligand
than individual receptor dimers. Note that this effect is weaker in
CheR mutants, but still existent.

Our analysis deals only with the early response to changes in ligand
concentration, since we have neglected methylation of receptors and
downstream phosphotransfer reactions.  It remains to integrate the
model for the early response developed here with a complete model such
as given in (Spiro et al. 1997) for later events.  It is of course
feasible to do this computationally, but given the complexity of the
association scheme for the formation of signaling teams shown in
Figure \ref{purescheme}, it may be difficult to extract qualitative
insights analytically. Some simplification exploiting the disparity in
time scales of the various processes will certainly be needed.

\medskip

\noindent
This work was supported by NIH Grant \#GM-29123 to
H. G. Othmer. We thank Sandy Parkinson for helpful discussions at
various stages of the model development.

Ames, P., C.~A. Studdert, R.~H. Reiser and J.~S. Parkinson. 2002. Collaborative 
signaling by mixed chemoreceptor teams
in Escherichia coli, {\em Proc. Natl. Acad. Sci. USA}  99: 7060-7065.

Barkai, N and Leibler, S. 1997. Robustness in simple biochemical networks.
{\em Nature (London)} 387:913-917.

Barnakov, A.~N, L.~A. Barnakova and G.~L. Hazelbauer. 1998. Comparison in vitro 
of a high- and a low-abundance chemoreceptor of {\it Escherichia Coli}: Similar kinase 
activation but different methyl-accepting activities, {\em J. Bacteriol.}  180:
6713--6718.

 Berg, H.~C. and R.~A. Anderson. 1973. Bacteria swim by rotating their flagellar 
filaments. {\em Nature} 245: 380-382.

Berg, H.~C. and D.~A. Brown. 1972. Chemotaxis in Escherichia coli analyzed 
by three-dimensional tracking. {\em Nature} 239:500-504.

Block, S.~M, J.~E Segall and H.~C. Berg. 1982. Impulse responses in bacterial 
chemotaxis. {\em Cell} 31:215-226.

Bornhorst, J.~A and J.~J. Falke. 2001. Evidence that both ligand binding and covalent
adaptation drive a two-state equilibrium in the aspartate receptor signaling complex 
{\em J. Gen. Physiol.}  118: 693-710.

Bourret, R.~B, K.~A Borkovich and M.~I. Simon. 1991. Signal transduction pathways 
involving protein phosphorylation in prokaryotes. {\em Annu. Rev. Biochem.} 
60: 401-41.

Bray, D, M.~D. Levin and C.~J. Morton-Firth. 1998. Receptor clustering as a 
cellular mechanism to control sensitivity. {\em Nature (London)}  393: 85-88.

Cluzel, P., M. Surette and S. Leibler. 2000. An ultrasensitive bacterial motor 
revealed by monitoring signaling proteins in single cells. {\em Science}  287:1652-1655.

Doedel, E.~J. 1981. AUTO: A program for the automatic bifurcation and
analysis of autonomous systems. {\it Cong. Num.} 30: 265-284.

Francis, N.~R., , M. N. Levit, T.~R. Shaikh, L.~A. Melanson, J.~B. Stock and
D.~J. DeRosier. 2002. Subunit organization in a soluble complex of Tar,
CheW, and CheA by electron microscopy, {\em J. Biol. Chem.}  277: 36755-36759.

Kim, C, M. Jackson, R. Lux and S. Khan. 2001. Determinants of chemotactic signal 
amplification in Escherichia coli. {\em J. Mol. Biol.}  307: 119-135.

Kim, S.~H, W. Wang and K.~K. Kim. 2002. Dynamic and clustering model of 
bacterial chemotaxis receptors: structural basis for signaling and high sensitivity.
{\em Proc. Natl. Acad. Sci. USA}  99: 11611-11615.

Levit, M.~N, Y. Liu and  J.~B. Stock. 1998. Stimulus response 
coupling in bacterial chemotaxis: receptor dimers in signalling arrays.
{\em Mol. Microbiol.}  30: 459-465.

Levit, M.~N. and J.~B.Stock. 2002. Receptor methylation controls the magnitude of
stimulus-response coupling in bacterial chemotaxis, {\em J. Biol. Chem.} 
277: 36760-36765.

Li, G. and  R.~M. Weis. 2000. Covalent modification regulates ligand binding to
receptor complexes in the chemosensory system of Escherichia coli.
{\em Cell}  100: 357-365.

 Macnab, R.~M., and M.~K. Ornston. 1977. Normal-to-curly flagellar transitions and their role in 
bacterial tumbling. Stabilization of an alternative quaternary structure by mechanical force. 
{\em J. Mol. Biol.} 112: 1-30.

Maddock, J.~R and L. Shapiro L. 1993. Polar location of the chemoreceptor complex in the
Escherichia coli cell. {\em Science}  259: 1717-1723.

Mello, B. A. and Y. Tu. 2003. Quantitative modeling of sensitivity in bacterial 
chemotaxis: The role of coupling among different chemoreceptor species.
{\em Proc. Natl. Acad. Sci. USA}  100: 8223-8228.

Morton-Firth, C.~J, T.~S Shimizu and D. Bray.
(1999 A free-energy-based stochastic simulation of the Tar receptor complex,
{\em J. Mol. Biol.}  286: 1059-1074.

Segall, J.~E, S.~M Block and H.~C. Berg. 1986. Temporal comparisons in bacterial 
chemotaxis. {\em Proc. Natl. Acad. Sci. USA}  83:8987-8991.

Spiro, P.~A, J.~S Parkinson and H.~G. Othmer. 1997. A model of excitation and 
adaptation in bacterial chemotaxis. {\em Proc. Natl. Acad. Sci. USA}  94:7263-7268.

Sourjik, V. and H.~C. Berg. 2002. Receptor sensitivity in bacterial chemotaxis.
{\em Proc. Natl. Acad. Sci. USA}  99:123-127.

Turner, L., W. Ryu and H.~C. Berg. 2000. Real-time imaging of fluorescent flagellar
filaments. {\em J. Bacteriol.} 182: 2793-2801.

\end{document}